\documentclass[journal,12pt,onecolumn, draftclsnofoot]{IEEEtran}
\usepackage[dvips]{graphicx}
\usepackage{multirow}
\usepackage{caption}
\usepackage{multicol,lipsum}
\usepackage{mathtools, cuted}
\usepackage{comment}
\usepackage{amsmath}
\usepackage{amsthm}
\usepackage[flushleft]{threeparttable}
\usepackage{array}
\usepackage{booktabs}
\newcommand{\PreserveBackslash}[1]{\let\temp=\\#1\let\\=\temp}
\newcolumntype{C}[1]{>{\PreserveBackslash\centering}p{#1}}
\newcolumntype{R}[1]{>{\PreserveBackslash\raggedleft}p{#1}}
\newcolumntype{L}[1]{>{\PreserveBackslash\raggedright}p{#1}}

\usepackage{bm}
\usepackage{graphicx,subfigure}
\usepackage{algorithmic, algorithm}
\usepackage{cite}
\usepackage{amsmath}
\usepackage{amssymb}
\usepackage{psfrag}
\usepackage{empheq}
\usepackage{latexsym, amsmath, subfigure, color, amsfonts, amssymb,graphicx}
\usepackage{wrapfig} 

\IEEEoverridecommandlockouts
\usepackage{tabularx}
\makeatletter
\def\hlinewd#1{%
\noalign{\ifnum0=`}\fi\hrule \@height #1 %
\futurelet\reserved@a\@xhline}
\makeatother
\begin{document}
\title{CNN-based Analog CSI Feedback in FDD MIMO-OFDM Systems}
\author{Mahdi Boloursaz Mashhadi, Qianqian Yang, and Deniz G\"{u}nd\"{u}z\\
\IEEEauthorblockA{Dept. of Electrical and Electronic Eng., Imperial College London, UK\\
Email: \{m.boloursaz-mashhadi, q.yang14, d.gunduz\}@imperial.ac.uk}}

\maketitle
\begin{abstract}
Massive multiple-input multiple-output (MIMO) systems require downlink channel state information (CSI) at the base station (BS) to better utilize the available spatial diversity and multiplexing gains. However, in a frequency division duplex (FDD) massive MIMO system, CSI feedback overhead degrades the overall spectral efficiency. Convolutional neural network (CNN)-based CSI feedback compression schemes has received a lot of attention recently due to significant improvements in compression efficiency; however, they still require reliable feedback links to convey the compressed CSI information to the BS. Instead, we propose here a CNN-based analog feedback scheme, called AnalogDeepCMC, which directly maps the downlink CSI to uplink channel input. Corresponding noisy channel outputs are used by another CNN to reconstruct the DL channel estimate. Not only the proposed  outperforms existing digital CSI feedback schemes in terms of the achievable downlink rate, but also simplifies the operation as it does not require explicit quantization, coding and modulation, and provides a low-latency alternative particularly in rapidly changing MIMO channels, where the CSI needs to be estimated and fed back periodically. 

\end{abstract}
\section{Introduction}
Massive multiple-input multiple-output (MIMO) systems are at the center of 5G and future wireless networks due to their ability to serve many users simultaneously with high spectral efficiency. A massive MIMO base station (BS) requires accurate and timely downlink channel state information (CSI) to achieve its promised performance gains. Although frequency division duplex (FDD)  operation mode in a massive MIMO setting is favourable due to its improved coverage and reduced interference, these benefits come at the price of increased CSI estimation and feedback overhead. In FDD massive MIMO, downlink CSI needs to be estimated by the user equipments (UEs) utilizing the pilots received from the BS and then fed back to the BS through the uplink channel. The CSI feedback rate is hence limited by the amount of uplink channel resources assigned to CSI feedback and the uplink channel quality. It becomes crucial to design an efficient CSI compression and feedback scheme that provides more accurate CSI at the BS while introducing a limited feedback overhead.

There are two main approaches in designing
the CSI feedback link, i.e., the \textit{digital} and \textit{analog} CSI schemes. Digital schemes, which have traditionally received more attention \cite{DigCSI1, DigCSI3, DigCSI4}, are based on Shannon’s source-channel separation theorem: CSI is first compressed into as few bits as possible and these bits are reliably fed back to the transimitter using a low-rate channel code. On the other hand, analog CSI follows a joint source-channel coding approach, and directly maps the downlink CSI to the uplink channel input in an unquantized and uncoded manner \cite{UQUC1, AnaCSI1, AnaCSI3}. The analog scheme simplifies the feedback operation and achieves the optimum mean-squared error in some cases (e.g., an additive white Gaussian noise (AWGN) feedback channel) while avoiding the quantization, coding and modulation delay.

Following the recent resurgence of machine learning, and more specifically deep learning (DL) techniques for physical layer communications \cite{MLintheAir}, DL-based MIMO CSI compression techniques have been shown to provide significant improvements over previous works utilizing compressive sensing and sparsifying transforms \cite{wen2018deep, Access1}. DL approaches use autoencoder architectures to compress the CSI, and most of them \cite{wen2018deep, Access1, DLCSI2, DLCSI3, DLCSI4, CSINETPlus} train the autoencoder assuming ideal feedback of the compressed CSI from the UE to the BS. However, the encoder output is fed back to the BS through the uplink channel, and needs to be encoded to overcome channel impairments (e.g. noise, fading, etc.) \cite{yang2019deep}.

While the previous works on DL-based CSI feedback focus on the digital CSI feedback scheme \cite{yang2019deep, shlezinger2019deep, lu2019bitlevel, FEDDEL}, here we design a DL-based analog feedback scheme taking into account the uplink channel explicitly. We use a fully convolutional autoencoder model to efficiently map the downlink CSI at the UE to the uplink channel inputs, and to reconstruct them at the BS. We train the model treating the uplink feedback channel as a non-trainable layer. We compare the analog and digital schemes based on the quality of the reconstructed CSI at the BS and the corresponding downlink rate achieved when the same amount of uplink channel resources is devoted to CSI feedback. The analog scheme improves the CSI accuracy and downlink rate without requiring the UL CSI at the UE for feedback transmission. If the feedback channel capacity is low due to noise, fading, etc., the digital scheme totally fails to deliver the CSI. This "threshold effect" degrades the downlink rate significantly. The analog scheme however, experiences graceful performance degradation and avoids the threshold effect.

\begin{figure*}[t!]
\centering
\includegraphics[scale=0.5]{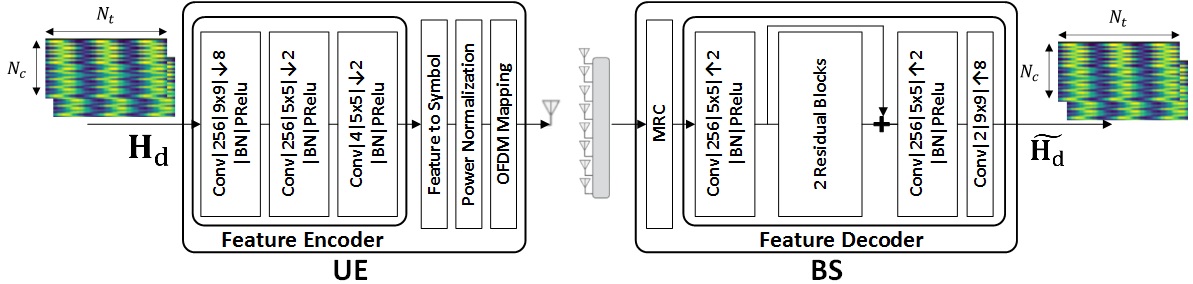}
\caption{The proposed model, AnalogDeepCMC, for downlink FDD MIMO analog CSI feedback.}
\label{fig2}
\end{figure*}

\section{System Model}\label{sec1}
Consider CSI feedback from a single-antenna user to a BS with $N_t$ antennas utilizing orthogonal frequency division multiplexing (OFDM) over $N_c$ subcarriers. Denote by $\mathbf{h}_{d}^{i} \in \mathbb{C}  ^{N_c \times 1}$ the downlink channel response from the $i$-th antenna of the BS to the user in the angular delay domain, and $\mathbf{H}_d\triangleq[\mathbf{h}_{d}^{1}  \mathbf{h}_{d}^{2}  \cdots  \mathbf{h}_{d}^{N_t}] \in \mathbb{C}  ^{N_c \times N_t}$. Similarly, we denote by $\mathbf{h}_{u}^{i} \in \mathbb{C}  ^{N_c \times 1}$ the uplink channel response from the user to the $i$-th antenna at the BS, and  $\mathbf{H}_u\triangleq[\mathbf{h}_{u}^{1} \mathbf{h}_{u}^{2} \cdots \mathbf{h}_{u}^{N_t}] \in \mathbb{C}  ^{N_c \times N_t}$. 

We assume that the downlink CSI $\mathbf{H}_d$ available at the UE is fed back to the BS over $N_f$ uplink OFDM subcariers devoted to CSI feedback picked uniformly at random, with $\rho\triangleq\frac{N_f}{N_c}$ denoted as the \textit{feedback overhead}. We denote the frequency domain uplink channel matrix  by $\mathbf{\hat{H}_u}$ obtained by column-wise FFT on $\mathbf{H}_u$. The feedback channel over the $j$-th uplink subcarrier  denoted by $\mathbf{\hat{h}}_{f}^{j} \in \mathbb{C}  ^{N_t \times 1}, j = 1, \cdots, N_f$, is obtained from the corresponding row of $\mathbf{\hat{H}_u}$, which specifies a SIMO channel with its output given by 
\small
\begin{align}\label{FBCh1}
    \mathbf{y}_j= \mathbf{\hat{h}}_{f}^{j} x_j + \mathbf{z}_j,
\end{align}
\normalsize
in which $\mathbf{y}_j \in \mathbb{C}  ^{N_t \times 1}$ is the received signal at the BS antennas, $x_j$ is the symbol fed back over the $j$-th subcarrier and $\mathbf{z}_j \in \mathbb{C}  ^{N_t \times 1}$ is the independent AWGN component.


With $N_f$ uplink sub-carriers dedicated for CSI feedback, a maximum rate of $C_{FB}=\sum_{j=1}^{N_f}\log_2(1+SNR_{FB}\|\mathbf{\hat{h}}_{f}^{j}\|^2)$ is possible for CSI feedback, where $SNR_{FB}$ is the signal to noise ratio (SNR) in the uplink channel. However note that $C_{FB}$ depends on the uplink channel state which is not known by the UE, hence, it will typically take a conservative approach and transmit back at a rate that can be decoded with high probability. Here, we use $C_{FB}$ as the feedback rate to get an upper bound on the performance of any digital CSI feedback scheme (which is practically very hard to achieve).

The goal of CSI feedback is to allow the BS to better focus its tranmit power towards the UE, in order to increase the average downlink transmission rate, which we use as the performance measure. Denoting the downlink CSI reconstructed at the BS and its frequency domain representation by $\widetilde{\mathbf{H}_d}$ and $\Hat{\widetilde{\mathbf{H}_d}}$, respectively, and asssuming that the BS employs conjugate beamforming, the average achievable downlink rate is
\small
\begin{align}\label{FBCh2}
    R_{DL}=\frac{1}{N_c} \sum_{i=1}^{N_c}\log_2 \left(1+SNR_{DL} \frac{\|\hat{{\mathbf{h}_{d}^{i}}} \cdot
\hat{\widetilde{\mathbf{h}_{d}^{i}}}\|^2}{\|\hat{\widetilde{\mathbf{h}_{d}^{i}}}\|^2} \right),
\end{align}
\normalsize
where $\hat{{\mathbf{h}_{d}^{i}}}$ and $\hat{\widetilde{\mathbf{h}_{d}^{i}}}$ are the $i$'th rows of $\Hat{{\mathbf{H}_d}}$ and $\Hat{\widetilde{\mathbf{H}_d}}$, and $SNR_{DL}$ is the downlink SNR. The more accurately the CSI can be reconstructed at the BS, the higher the inner product in (\ref{FBCh2}), and the higher the downlink rate. 

\begin{figure}[t!]
\centering
\includegraphics[scale=0.8]{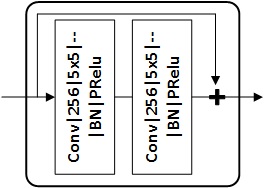}
\caption{The residual block Model.}
\label{fig3}
\end{figure}

\section{CSI Feedback Schemes}\label{section:2}

In this section, we present our models for analog and digital CSI feedback schemes. 

\subsection{Digital CSI Feedback}
 We employ the DeepCMC architecture proposed in \cite{yang2019deep}, which consists of a convolutional encoder at UE that encodes $\mathbf{H}_d$ to a variable-length bit stream, and a convolutional decoder at BS to reconstruct $\mathbf{H}_d$ from the received bit stream. The detailed description of the encoder and decoder can be referred to \cite{yang2019deep}.  The resulting bit stream of the encoder is channel encoded and modulated. The modulated signals are transmitted over the feedback OFDM subcarriers according to (\ref{FBCh1}). 
The received signal at BS then passes through demodulation, channel decoding, to give the bit stream input to the decoder. The encoder and the decoder are trained jointly according to the loss function presented in \cite{yang2019deep} with a parameter, $\lambda$, governing the tradeoff between reconstruction quality and the feedback rate. 


\begin{figure}[t!]
\centering
\includegraphics[scale=1]{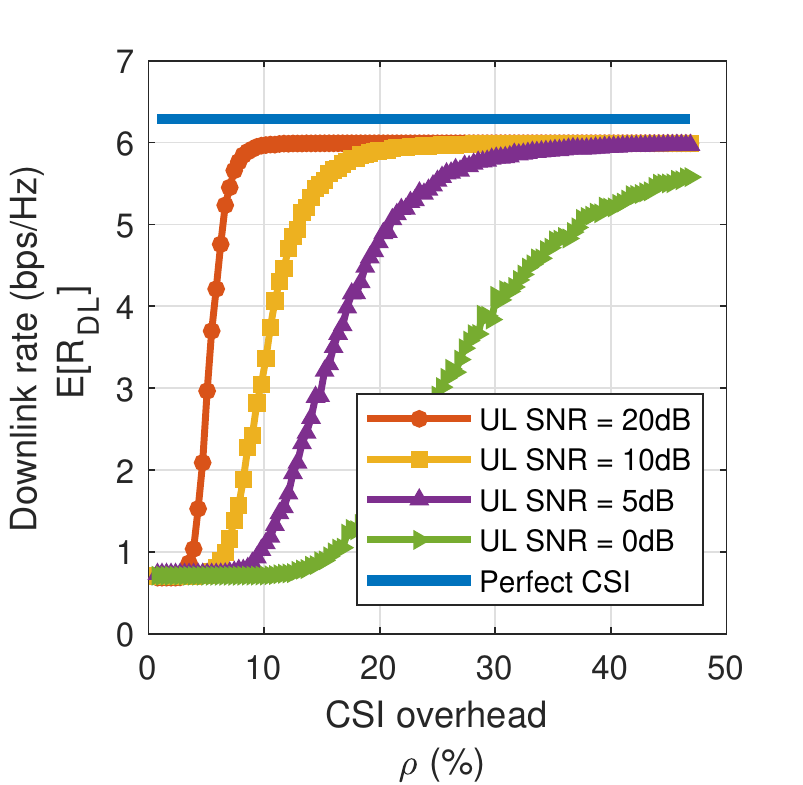}
\caption{Average downlink rate vs. $\rho$ for digital CSI feedback scheme.}
\label{curve1}
\end{figure}

\subsection{Analog CSI Feedback}
Fig.~\ref{fig2} depicts the architecture of our proposed analog CSI feedback model, Analog-DeepCMC. ``Conv$|256|$ $9\times9| \downarrow4|$BN$|$PRelu'' represents a convolutional layer with 256 $9\times9$ kernels followed by downsampling by a factor of 4, batch normalization and PRelu activation. The ``SAME'' padding technique is used with all convolutions. ``+'' denotes simple element-wise addition. Fig.~\ref{fig3} illustrates the architecture for each residual block where ``$|--$'' means the corresponding convolution output is not downsampled.

The UE applies a CNN-based feature encoder composed of three convolutional layers which outputs real-valued features. Each pair of these real numbers are then grouped to form a complex-valued symbol, which are subsequently normalized to ensure the input power constraint over the feedback channel is met. These normalized symbols are then directly mapped into the corresponding subcarriers, and transmitted over the CSI feedback channel. The BS then performs maximum ratio combining and feature decoding to reconstruct the original CSI matrix $\mathbf{H}_d$. Note that for different CSI overhead values $\rho$, the downsampling factors and the number of features should vary in Fig.~\ref{fig2}.

Unlike the digital scheme, the analog scheme is trained taking into account all the blocks in the model, including the feedback channel and MRC. The CSI feedback channel in (\ref{FBCh1}) and the MRC block both follow simple linear, differentiable models, and are accommodated as untrainable layers in the autoencoder architecture to improve the end-to-end performance without requiring explicit uplink CSI at the UE side.

\begin{figure}[t!]
\centering
\includegraphics[scale=1]{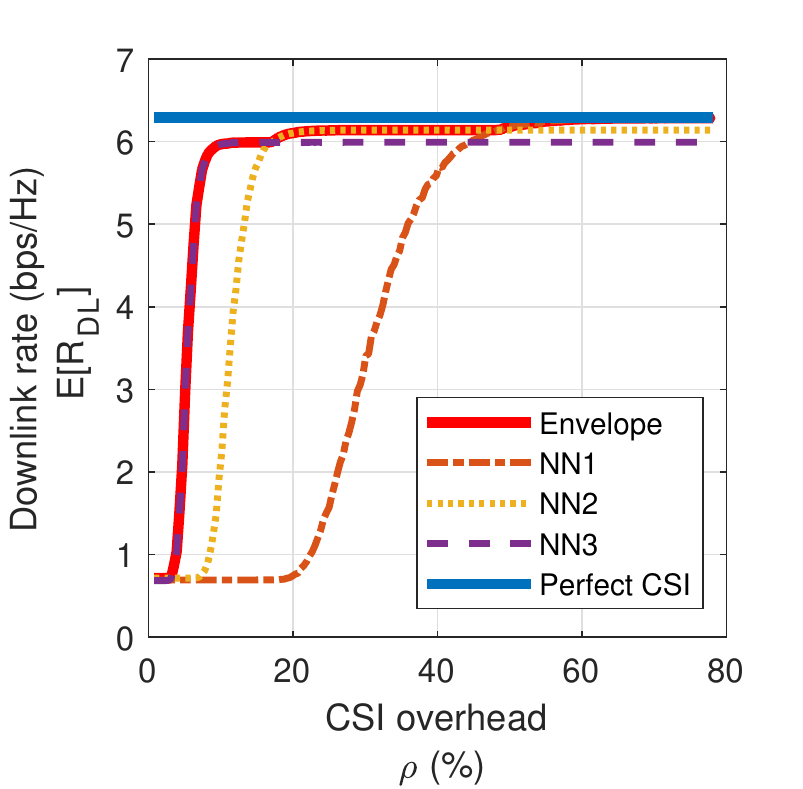}
\caption{Digital CSI performace for uplink SNR 20dB.}
\label{curve2}
\end{figure}

\section{Simulations}\label{sec4}

We use the COST 2100 channel model \cite{COST2100} to generate sample uplink/downlink channel matrices for training and testing. We consider the indoor picocellular scenario at 5.3 GHz, where the BS is equipped with a uniform linear array (ULA) of dipole antennas and positioned at the center of a $20\mathrm{m} \times 20\mathrm{m}$ square area, and any other parameter follows its default setting in \cite{COST2100}. The number of training and testing samples are 80000 and 20000, respectively, and the batch size is 100. We train our models for $N_c=256$ and $N_t=32$.

Fig. \ref{curve1} depicts the average downlink rate for different values of CSI overhead $\rho$ and uplink SNR using the digital scheme, where we recall that $\rho\triangleq\frac{N_f}{N_c}$, and $N_f$ is the number of subcarriers devoted to CSI feedback. For comparison, the average rate with perfect CSI available at the BS is also provided as the upper bound. In this case, if the size of encoder's output exceeds $C_{FB}$, the feedback channel fails to deliver the CSI. Without downlink CSI, the BS has to use the average CSI (that corresponds only to the line of sight (LOS) multi-path component of fading) for beamforming, resulting in the minimum rate. Due to this, the rate curves all show a threshold behaviour. If the CSI overhead decreases below a threshold, failures occur very frequently resulting in the minimal downlink rate. For overhead values well above the threshold, failures become rare and the autoencoder can reconstruct CSI at the distortion that it has been trained for, hence achieving an acceptable downlink rate. According to this figure, as the uplink SNR decreases, the performance threshold increases.

 \begin{figure}[t!]
\centering
\includegraphics[scale=1]{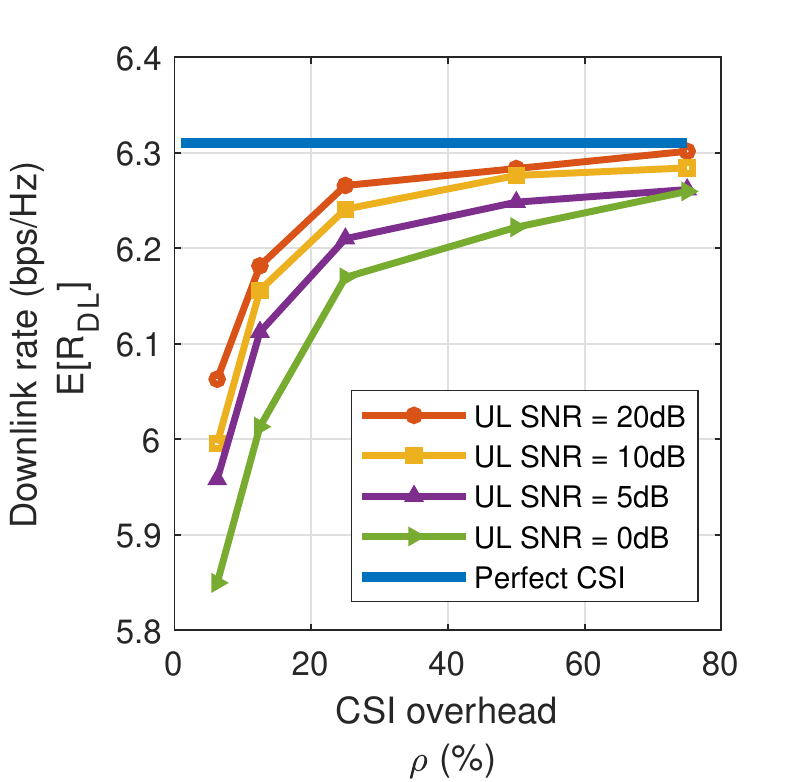}
\caption{Average downlink rate vs. $\rho$ for analog CSI feedback scheme.}
\label{curve3}
\end{figure}

As mentioned previously, the autoencoder model for the digital scheme can be trained for different reconstruction qualities by changing the rate-distortion weight parameter $\lambda$. 
Networks trained for different reconstruction qualities result in different threshold behaviours. A network trained for better reconstruction quality results in an increased performance threshold but achieves a better downlink rate for overhead values above the threshold. Fig. \ref{curve2} shows the average downlink rate curves for three different NNs trained for different $\lambda$ values where the uplink SNR is 20dBs. For improved performance, the UE can save the trained parameter values for three networks and use the network with the best reconstruction quality for each value of overhead $\rho$. This gives the envelope curve for the overall performance of the digital CSI approach, which we later compare with the analog scheme.

Fig. \ref{curve3} depicts the average downlink  rate achieved by AnalogDeepCMC. The curves for different SNR values in Fig. \ref{curve3} correspond to NN models trained for the corresponding uplink SNR. Simulation results show that for uplink SNR values in the range 0-20 dB, the performance still remains acceptable which is unlike the digital case which may result in a failure to decreased uplink SNR (threshold effect). 

\begin{figure}[t!]
\centering
\includegraphics[scale=1]{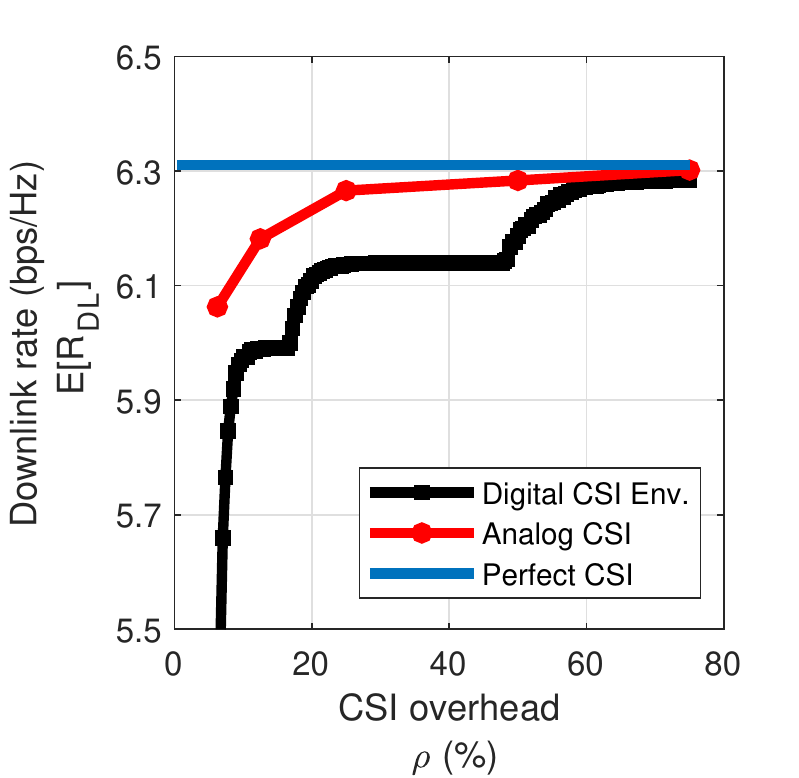}
\caption{Average downlink rate for uplink SNR 20dB.}
\label{curve4}
\end{figure}

Fig. \ref{curve4} compares the performance of the digital and analog CSI schemes where the uplink SNR is 20dBs. While both curves approach the perfect CSI bound as the feedback overhead increases, the analog curve provides up to $0.1928$ bps/Hz (almost $3.1\%$ of the perfect CSI upper bound) improvement in the achieved downlink rate. The improvement becomes more significant as the uplink SNR decreases (e.g. $1.2161 \quad \mathrm{bps/Hz} = 19.3\%$  for uplink SNR 10 dBs). Note that we have been very generous to the digital scheme by assuming capacity-achieving channel codes. Note that, for the setting considered here ($N_c=256$), $\rho=20\%$ would correspond to a channel code of length 51 symbols, in which case the code rates with reasonable reliability are significantly below the capacity \cite{polyanskiy2010channel}. 

\section{Conclusion}\label{sec5}
We proposed CNN-based CSI feedback schemes to efficiently feedback the downlink CSI in a FDD massive MIMO scenario. We compared the performance of digital and analog feedback schemes considering unknown and time-varying uplink feedback channel. Simulations showed that the analog scheme not only provides a low-latency CSI feedback scheme avoiding the separate quantization, coding and modulation processes, but also improves the end-to-end CSI reconstruction quality, and hence, the achievable downlink rate.

\bibliographystyle{IEEEtran}
\bibliography{refs}
\end{document}